# Reconstruct the Hierarchical Structure in a Complex Network

Huijie Yang[1*], Fangcui Zhao[2], Wenxu Wang[1], Tao Zhou[1] and Binghong Wang[1]

[1] Department of Modern Physics and Nonlinear Science Center, University of Science and Technology of China, Hefei Anhui 230026, China

[2] College of Life Science and Bioengineering, Beijing University of Technology, Beijing 100022, China

## Abstract

A number of recent works have concentrated on a few statistical properties of complex networks, such as the clustering, the right-skewed degree distribution and the community, which are common to many real world networks. In this paper, we address the hierarchy property sharing among a large amount of networks. Based upon the eigenvector centrality (EC) measure, a method is proposed to reconstruct the hierarchical structure of a complex network. It is tested on the Santa Fe Institute collaboration network, whose structure is well known. We also apply it to a Mathematicians' collaboration network and the protein interaction network of Yeast. The method can detect significantly hierarchical structures in these networks.

**PACS (numbers)**: 89.75.2k, 89.20.Hh, 05.65.1b

---

[*] Corresponding author, E-mail: huijieyangn@eyou.com

Very diverse systems in different research fields can be described as complex networks, i.e., connecting the nodes together by the edges with nontrivial topological structures [1]. Detailed works have been focused on several distinctive statistical properties sharing among a large amount of real world networks, to cite examples, the clustering effect [2,3], the right-skewed degree distribution [4,5,6], and the community structure ([7] and for a recent review see Ref. [8]), etc. In this paper we consider another property sharing among many networks, the hierarchical structure of a complex network.

Hierarchy, as one common feature for many real world networks, attracts special attentions in recent years [9-12]. In a network, there are usually some groups of nodes, where the nodes in each group are highly interconnected with each other, while there are few or no links between the groups. These groups can induce high degree of clustering, which can be measured with the connectivity probability for a pair of the neighbors of one node. This property coexists usually with the right-skewed degree distributions. The coexistence of these two properties tells us that the groups should combine into each other in a hierarchical manner. Hierarchy is one of the key aspects of a theoretical model [9,13] to capture the statistical characteristics of a large number of real networks, including the metabolic [14,15], the protein interaction [16,17] and some social networks [18-20].

In literature, several concepts are proposed to measure the hierarchy in a network, such as the hierarchical path [10], the scaling law for the clustering coefficients of the nodes [9], the hierarchical components/degree [11], etc. These measures can tell us the existence and the extent of hierarchy in a network. We address herein another problem, that is, how to reconstruct the hierarchical structure in a network. If the nodes are positioned randomly in the topological map of a complex network, it is hard for us to find the hierarchical structure, even for a regular network. For a real network, there are some shortcuts disturb the simple structure. Reconstructing the hierarchical structure is clearly a nontrivial task.

Consider a network represented with the adjacent matrix $A$, the element $A_{ij}$ is $1$ and $0$ if the nodes $i$ and $j$ are connected and disconnected, respectively. The eigenvector centrality (EC) [21,22] is employed as the proxy of importance of each node. Denote the eigenvector corresponding to the principal eigenvalue of this adjacency matrix with $e_{principal}$, the eigenvector centrality of the $i$'s node is the $i$'s component of $e_{principal}$, i.e., $EC(i) = N_0 \cdot |e_{principal}(i)|^2$, where $N_0$ is the number of the nodes. This measure simulates a mechanism in which each node affects all of its neighbors simultaneously.

For a simple network containing only one group, the intra-group hierarchical structure can be detected by calculating EC values for all the nodes. Fig.1 and Fig.2 illustrate the EC values for the nodes in a Cayley-tree network. This measure can find the treelike structure exactly, based upon which the schematic illustration can be reconstructed in detail. To represent the hierarchical structure, we can catalogue the nodes in one group into several levels according to their EC values. The edges between the neighboring levels and between the nodes in a same level can be reserved as the initial structure, while the other edges can be regarded as shortcuts.

But this procedure may induce fatal mistakes in dealing with complex networks. Consider a general condition where several subordinate groups are connected loosely with a central group. The EC values obtained from the original adjacent matrix reflect mainly the structure of the central group. The EC values of the nodes in the subordinate groups will tend to vanish, because there are few nonzero elements between the central group and the subordinate groups in the corresponding adjacent matrix. Hence, the subordinate groups will be lost if we detect the key nodes with the EC values of the original adjacent matrix only. To resolve this problem, we should remove the found key nodes from the original adjacent matrix and calculate the EC values of the left nodes to find the possibly lost key nodes. Hence a proper procedure can be illustrated as follows.

Calculate the EC values for all the nodes from the original adjacent matrix $A_1$. Introducing a critical EC value, $EC_{crit} = \gamma \cdot \max(EC)$, the nodes whose EC values are larger than $EC_{crit}$ can be regarded as the key nodes. We can adjust the parameter $\gamma$ in the range of $[0,1]$. Then remove the found key nodes from the initial adjacent matrix and obtain a new adjacent matrix $A_2$.

Iteration of this step leads to some sets of keys nodes as, $S_m(n)|m = 0,1,2,3,\cdots,M$ and the corresponding adjacent matrix, $A_m|m = 1,2,3,\cdots,M$. Here, $S_m(n)$ is the set containing $n$ key nodes found at the $m$'th step. Each set of key nodes can form a backbone in the corresponding group. Catalogue the nodes in each set into several levels we can obtain the intra-group structure. Generally, we can sort the key nodes in each group in an ascending way.

To terminate the detecting procedure we should design a criterion. Define the average number of edges per key node, and the average number of edges per left node as,

$$D_{key}(m) = \frac{K_{key}(m)}{N_{key}(m)},$$
$$D_{left}(m) = \frac{K_0 - K_{key}(m)}{N_0 - N_{key}(m)},$$
(1)

where $K_{key}(m)$ and $N_{key}(m)$ are the edges between the key nodes and the total number of the key nodes found up to the $m$'th step, respectively. The terminate criterion can be designed as $D_{left}^c$. Once the average edges per left node $D_{left}(m)$ decreases to $D_{left}(m) \prec D_{left}^c$, the procedure is terminated.

By the above procedure we can find the global hierarchy and the intra-group hierarchy simultaneously. As an example, we consider the largest component of the Santa Fe Institute collaboration network [23]. It contains $118$ scientists working in four divisions of disciplinary. There are $192$ edges between these nodes. The first set of key nodes is the four nodes with red colors (as shown in Fig.3). They are the central nodes connecting the left groups together. The solid circles with different colors are the key nodes in the subgroups. The open circles are the nodes omitted in the detection for their being nonessential in the subgroups. These omitted nodes are basically in the skirts of the subgroups and the degrees of some nodes are just $1$. The adjacent matrix for the key nodes in Fig.4 shows that there are seldom connections between the subgroups. The central group (node 1,2,3 and 4) connects the subgroups together. In each subgroup there are one or two central nodes connecting the other intra-group nodes tightly. A significant hierarchy property can be detected. The average edge number in the initial network is about $1.63$, the value of the criterion is chosen as, $D_{key}^c = 1.5$. The parameter $\gamma$ is assigned the value of $0.1$.

The Erdos Number Project homepage [24] provide a mathematicians' collaboration network containing 6927 nodes connected with 12864 edges. The parameters $D_{key}^c$ and $\gamma$ are assigned the values of $1.5$ and $0.1$, respectively. In Fig.5 we can find that the nodes from 1 to 200 condensed into several central groups, the intra-group nodes are tightly connected and there are many connections between these central groups. These central groups connected all the subgroups together. In each subgroup there are some central nodes connecting the other intra-group nodes tightly. A significant hierarchy property can be detected. The most sub-groups are very small. Fig.6 presents the average numbers of edges per key node and per left node. It shows that all the groups form globally a tight-connected backbone of the initial network.

In the protein interaction network of Yeast [25] some key nodes in each group connected tightly to form a core. Though significant intra-group structures can be found, there is not a global hierarchical structure, as shown in Fig.7 and Fig.8

In summary, by means of the EC measure we propose a new method to reconstruct the hierarchy property in a complex network. The hierarchical structure in two real world collaboration networks can be detected effectively. Omitting the skirt nodes, we can draw a hierarchy-structured backbone of a network. The skirt nodes can be regarded as perturbations.

Besides of the ability to detect effectively the hierarchy property, this method can also provide useful information in identifying communities in a large network. The low level subgroups should be corresponding to the cores of the communities. It provides the importance of each node in each core in detail.

We acknowledge partial support from the National Science Foundation of China (NSFC) under Grant No.70471033, No.10472116 and No.70271070. It is also supported by the Specialized Research Fund for the Doctoral Program of Higher Education (SRFD No.20020358009). One of the authors (H. Yang) would like to thank Prof. Y. Zhuo, Prof. J. Gu in China Institute of Atomic Energy and Prof. S. Yan in Beijing Normal University for stimulating discussions.

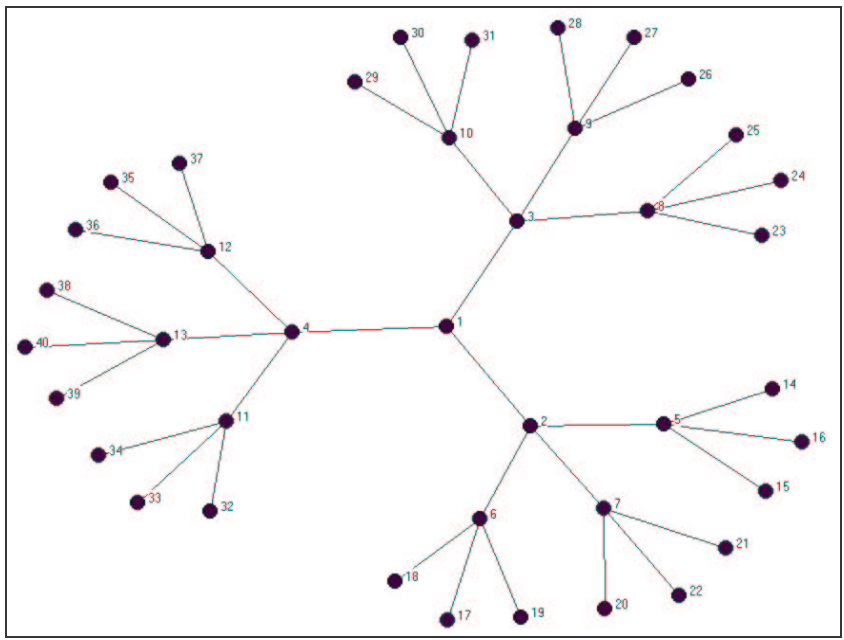

**Fig.1** A Cayley tree with the branching factor $z = 3$ and the level of the leaves $l = 3$.

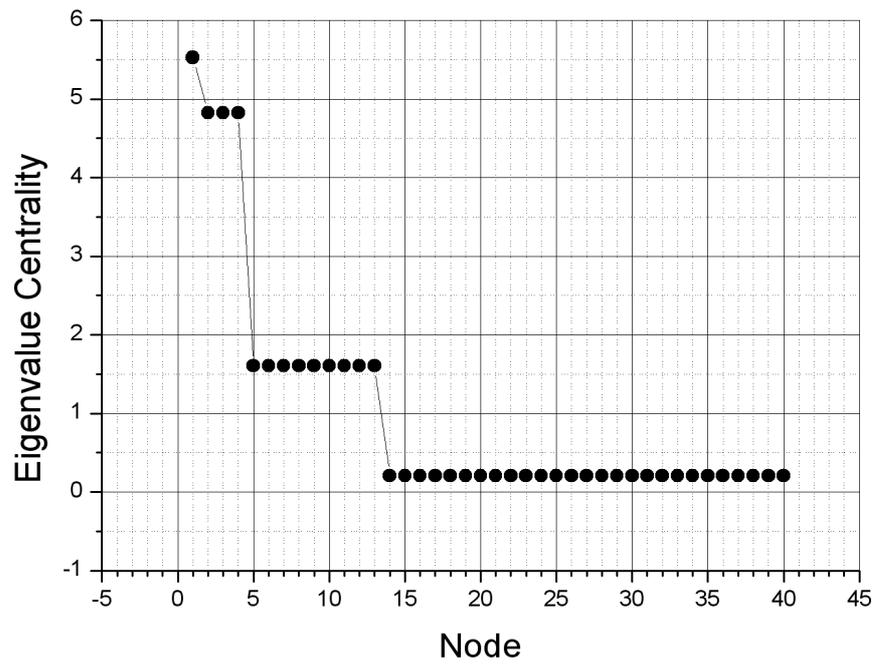

**Fig.2** The hierarchical structure of the Cayley tree presented in Fig.1. The measure can find the levels exactly, based upon which we can reconstruct the structure in fig.1 from a randomly positioned topological map.

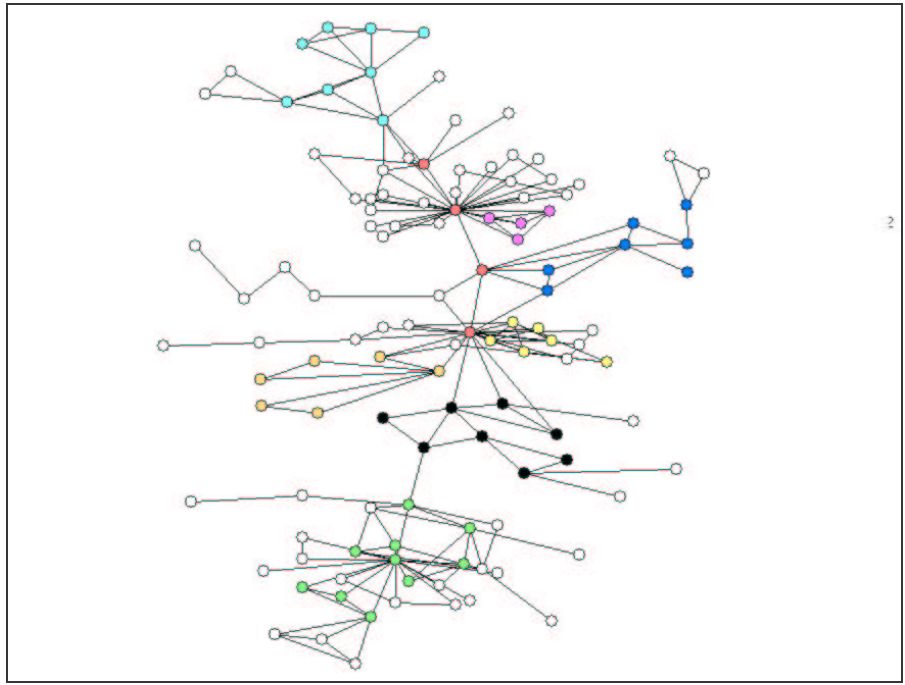

**Fig.3** The largest component of the Santa Fe Institute collaboration network. The first set of key nodes is the four nodes with red colors. They are the central nodes connecting the left groups together. The solid circles with different colors are the key nodes in the subgroups. The open circles are the nodes omitted in the detection for their being basically in the skirts of the subgroups. The parameters $D_{left}^{c}$ and $\gamma$ are assigned the values of $1.5$ and $0.1$, respectively.

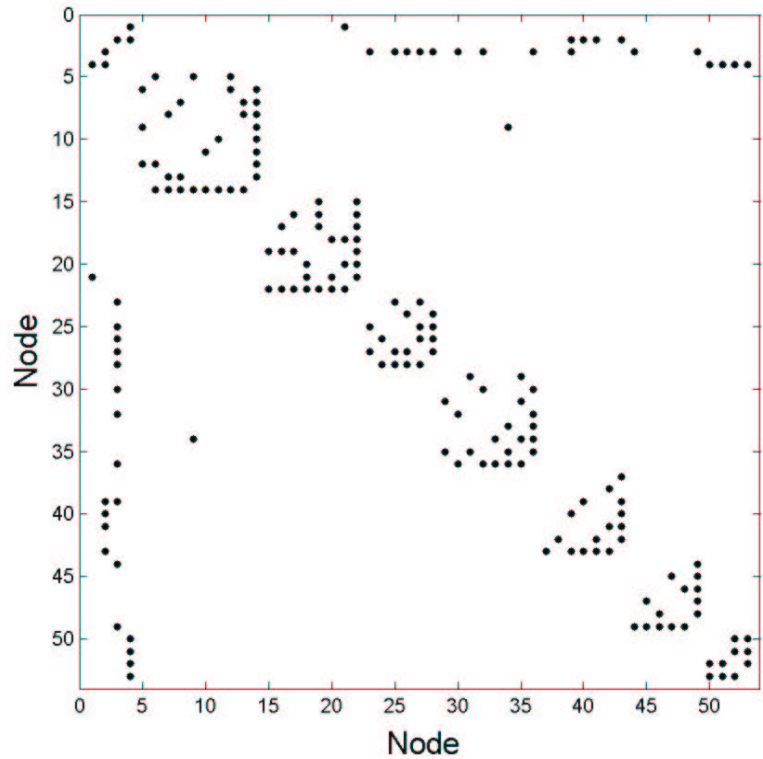

**Fig.4** The adjacent matrix for the key nodes found in the largest component of the Santa Fe Institute collaboration network. There are seldom connections between the subgroups. The central group (node 1,2,3 and 4) connects the subgroups together. In each subgroup there are one or two central nodes connecting the other intra-group nodes tightly. A significant hierarchy property can be detected. The parameters $D_{left}^{c}$ and $\gamma$ are assigned the values of $1.5$ and $0.1$, respectively.

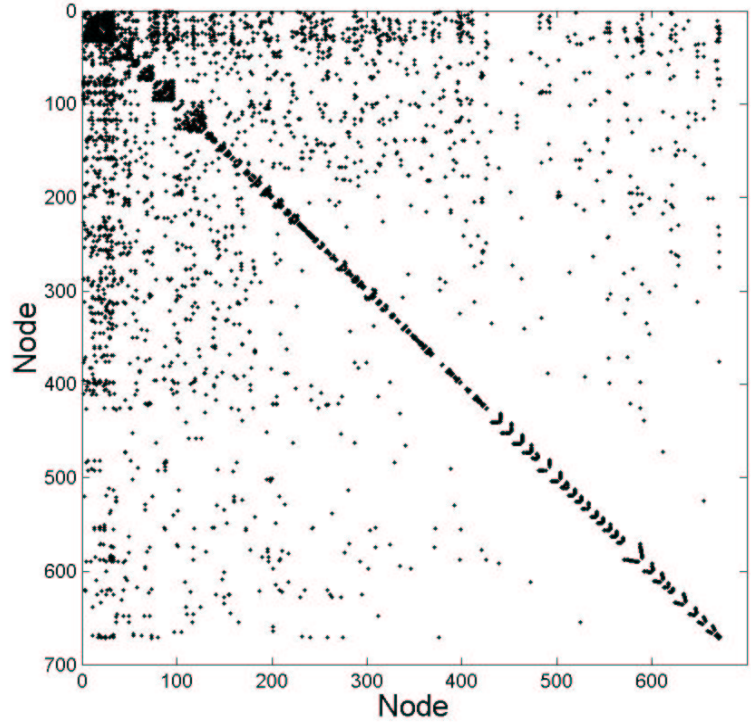

**Fig.5** The adjacent matrix for the key nodes in the mathematicians' collaboration network. The nodes from 1 to 200 condensed into several central groups, the intra-group nodes are tightly connected and there are many inter-group connections between these central groups. These central groups connected all the subgroups together. In each subgroup there are some central nodes connecting the other intra-group nodes tightly. A significant hierarchy property can be detected. The most sub-groups are very small. The parameters $D_{left}^{c}$ and $\gamma$ are assigned the values of $1.5$ and $0.1$, respectively.

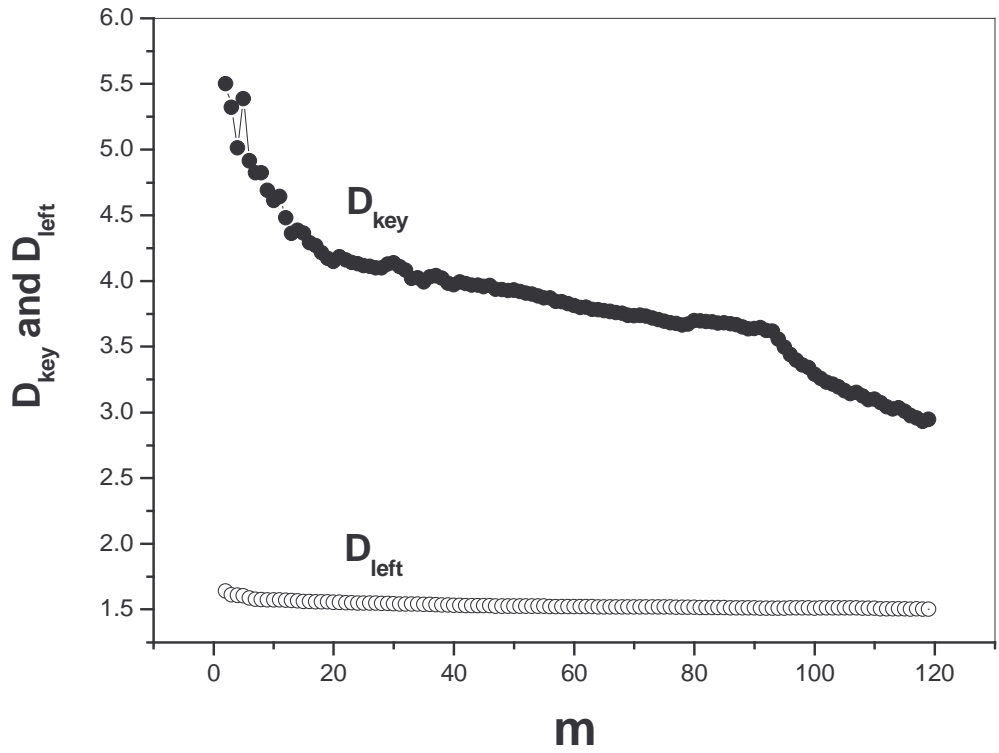

**Fig.6** The average numbers of edges per key node and per left node in the mathematicians' collaboration network. It shows that all the groups form globally a tight-connected backbone of the initial network.

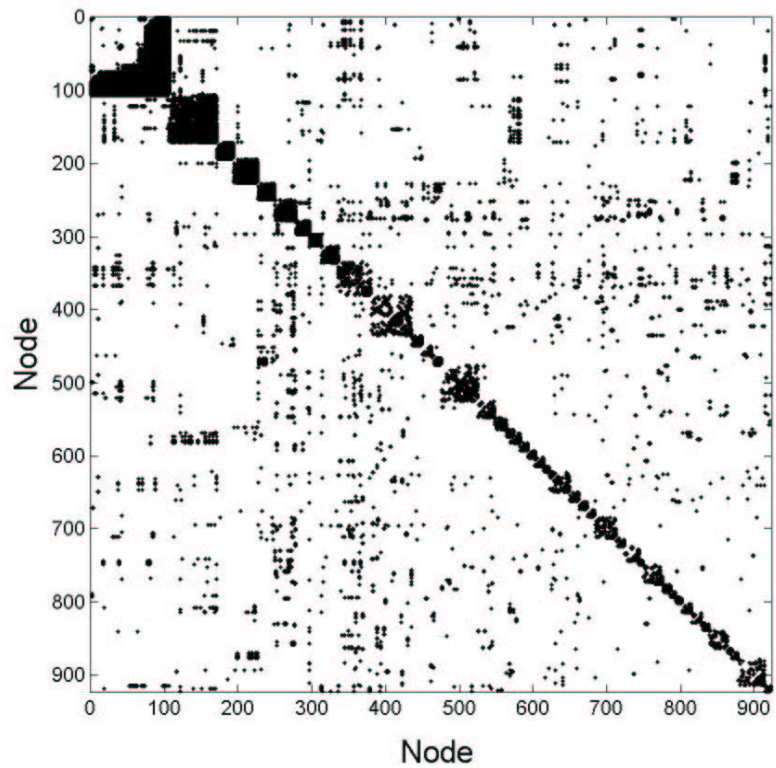

**Fig.7** The protein interaction network of Yeast. Some key nodes in each group connected tightly to form a core. Though significant intra-group structures can be found, there is not a global hierarchical structure. The parameters $D_{left}^{c}$ and $\gamma$ are assigned the values of $2.2$ and $0.1$, respectively.

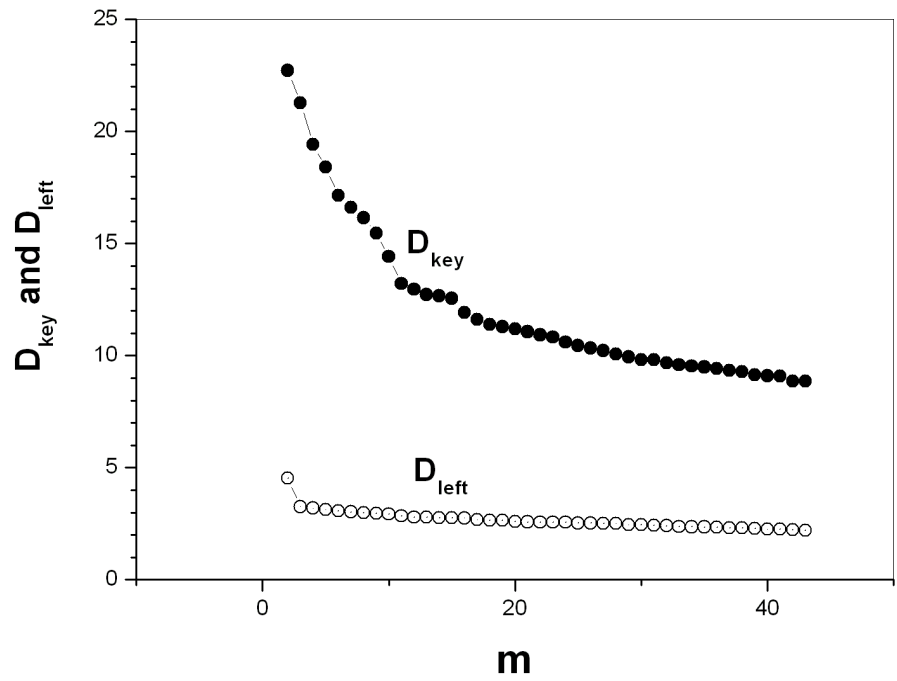

**Fig.8** The average numbers of edges per key node and per left node in the protein interaction network of Yeast. It shows that all the groups form globally a tight-connected backbone of the initial network.